\documentclass[a4paper,11pt]{article}
\pdfoutput=1 % if you are submitting a pdflatex (i.e. if you have
\usepackage{jheppub}

\usepackage{graphicx, epstopdf}

\usepackage[utf8]{inputenc}

\usepackage{amsmath, amsthm, amsfonts, amssymb}
\usepackage{mathtools, mathabx, bm, bbm, scalerel, xfrac}
\usepackage{empheq, physics, slashed, cancel}
\usepackage{booktabs, makecell, siunitx, adjustbox}

\usepackage{booktabs,tabularx}
\usepackage{listings}
\lstdefinestyle{cppapi}{
  language=C++,
  basicstyle=\ttfamily\small,
  columns=fullflexible,
  keepspaces=true,
  breaklines=true,
  showstringspaces=false
}
\newcommand{\code}[1]{\texttt{\detokenize{#1}}}

\usepackage{hyperref}
\usepackage{color}
\usepackage{xspace} % decides whether to insert a space or not

\allowdisplaybreaks

\title{Solution of Canonical Differential Equations for Integrals on Arbitrary Geometries}

\author{Micha\l{} Czakon and}
\author{Lorenzo Tancredi}
\affiliation{Institut f\"ur Theoretische Teilchenphysik und Kosmologie, RWTH Aachen University,\\ D-52056 Aachen, Germany}
\affiliation{Technical University of Munich, TUM School of Natural Sciences, Physics Department,\\ James-Franck-Straße 1, 85748 Garching, Germany}
\emailAdd{mczakon@physik.rwth-aachen.de}
\emailAdd{lorenzo.tancredi@tum.de}

\abstract{
A highly successful approach to computing multi-loop scattering amplitudes is to reduce the Feynman integrals that arise to a smaller set of master integrals using integration-by-parts identities. These dimensionally-regulated master integrals can often be determined by solving a system of first-order partial differential equations with respect to 
masses and external invariants. 
The application of this method to large classes of problems
became much more streamlined thanks to the introduction
of $\epsilon$-factorized canonical forms.
There is increasing evidence that a canonical form can always be achieved, 
although the required transformation may involve transcendental functions related to the periods of geometrical objects such
as elliptic curves or Calabi-Yau manifolds.
 Until now, obtaining numerical values for the master integrals in such cases has been difficult in practice, also due to the lack of closed-form expressions for the transcendental functions involved.
We show that this obstruction is only apparent. Since the original master integrals satisfy linear differential equations with rational coefficients, any functions appearing in the transformation to a canonical basis satisfy, by construction, rational differential equations as well. 
By solving these auxiliary equations, the numerical evaluation of the canonical system reduces to solving an enlarged rational system. We implement this strategy in a C\texttt{++} package and apply it to the two-loop master integrals that enter di-jet and $\gamma$+jet hadro-production via a heavy-quark loop.
}
\preprint{P3H-26-047,TTK-26-18,TUM-HEP-1607/26}

\begin{document}
\maketitle
\flushbottom

\section{Introduction}

Precise theoretical predictions for scattering processes are a central ingredient of collider phenomenology. 
As experimental measurements become more accurate, the required perturbative calculations involve 
amplitudes with more loops, more scales and, importantly, massive internal and external particles. 
Given the impressive recent progress towards general algorithms to reorganize and subtract infrared divergences
and define finite physical observables to next-to-next-to-leading-order (NNLO) and beyond,
the evaluation of multiloop multiscale Feynman integrals is often the main bottleneck in these calculations.

A standard strategy is to reduce all integrals contributing to a given amplitude to a finite set of master integrals by means of integration-by-parts identities~\cite{Tkachov:1981wb, Chetyrkin:1981qh}. To calculate the resulting master integrals 
many methods have been developed, and one of the most powerful ones involves
the derivation and solution of systems of linear rational differential equations in the kinematic invariants and masses~\cite{Kotikov:1990kg,Bern:1993kr,Remiddi:1997ny,Gehrmann:1999as}. 
An important step forward in the application of this method to increasingly complex problems has been 
the introduction of so-called canonical $\epsilon$-factorized differential equations~\cite{Henn:2013pwa}. 
In the polylogarithmic case, the canonical form exposes the iterated-integral structure of the answer by separating 
the dependence on the dimensional regulator from the kinematic dependence. In its most standard definition,
a canonical form implies a connection matrix whose entries are differentials of logarithms, or in short, dlog forms. 

It has been clear for a long time, however, that multi-scale, multiloop Feynman integrals, especially when massive virtual particles are involved, often involve mathematical structures beyond simple dlog forms. 
In particular, many cases are known of Feynman integrals involving elliptic curves, Calabi-Yau (CY) manifolds and higher-genus curves, 
and there are currently no guarantees that more general geometries will not appear.\footnote{For a comprehensive study at the two-loop order see~\cite{Bargiela:2025vwl}.} 
Recently, substantial progress has been achieved in understanding how the concept of canonical basis can be generalized to
these cases~\cite{Adams:2016xah,Frellesvig:2021hkr,Pogel:2022ken,Dlapa:2022wdu,Gorges:2023zgv,Duhr:2025lbz,Chen:2025hzq,Chaubey:2025adn,e-collaboration:2025frv,Bree:2025tug,Duhr:2025xyy,Yang:2025ofz}, and a much clearer 
picture has started to emerge. In particular, the concept of leading singularity~\cite{Cachazo:2008vp} can be naturally
extended beyond logarithmic amplitudes~\cite{Bourjaily:2020hjv, Bourjaily:2021vyj,Forner:2026vby}, to 
show that integrals with unit leading
singularities in this generalized sense satisfy canonical epsilon-factorized differential equations even on
large classes of geometries~\cite{Forner:2026vby}.
The price to pay to attain such a canonical form is that the corresponding connection matrix is no longer necessarily rational or algebraic, as in the polylogarithmic case: its entries will in general contain transcendental functions associated with the periods of the underlying geometry.

Thanks to this progress, the practical question then becomes  how to extract numerical values from such systems. 
One obvious possibility is to construct local series expansions, either for the Feynman integrals or for the whole scattering amplitudes, and analytically continue them through phase space. This strategy has been applied
 very successfully for problems involving a small number of scales, but it is unclear how it would perform for multiscale
problems, as those inherently associated with the production of three or more particles in high-energy collisions.
A different strategy is to integrate the differential equations directly along paths in the space of kinematic variables.
 The latter approach is especially natural for linear systems with rational coefficients,
  for which robust numerical algorithms are available.
In fact, the direct numerical solution of differential equations for master integrals satisfying rational differential equations
has been pursued for more than two decades
and first calculations of complete virtual amplitudes using this method appeared in Refs.~\cite{Boughezal:2007ny,Czakon:2008zk}. 
In practice, those calculations relied on a solver written in Fortran. 
The transition to C\texttt{++} was an important step forward~\cite{Czakon:2020vql}, although the applications considered were still restricted to two-to-two scattering~\cite{Czakon:2021yub}. Recently, the method has been applied to two-to-three scattering~\cite{PetitRosas:2025xhm,Badger:2025ljy}, showing that problems with five kinematic variables can be solved efficiently and one can outperform approaches based on the dynamical generation and solution of series expansions~\cite{Hidding:2020ytt,Prisco:2025wqs}. 

Refs.~\cite{PetitRosas:2025xhm,Badger:2025ljy} also addressed, for the first time in this framework, the presence of square roots in the differential-equation system. The difficulties encountered there illustrate the challenge posed by canonical differential equations with more general algebraic or transcendental entries.
In fact, at first sight, canonical systems with algebraic and transcendental entries (including elliptic or CY periods)
seem to fall outside this framework, since evaluating the special functions appearing in the connection matrix is not straightforward if closed-form expressions for the latter are not available.
The main point of this paper is to demonstrate explicitly that this obstruction is only apparent. 
The non-rational functions appearing in a canonical transformation are not arbitrary functions: they arise from a system of Feynman integrals which, before the transformation to canonical form, satisfy linear differential equations with rational coefficients. Hence, all algebraic and transcendental building blocks that appear in the connection matrix can themselves be defined via the systems of  rational differential equations that they satisfy. 
By adjoining these auxiliary equations to the canonical system, one obtains an enlarged rational system that can be solved numerically with the same methods, without ever evaluating the special functions in closed form.
In this sense, one could say that canonical bases beyond the polylogarithmic case provide an efficient way to package the rational
information contained in the original differential equations into a minimal form.

Interest in numerical solutions of canonical differential equations has also grown recently. For instance, Ref.~\cite{Baur:2026zlw} presents a package that transforms iterated integrals into a differential-equation system and solves it numerically. Since iterated integrals are the formal solution of canonical differential equations, that work is conceptually related to ours. Ref.~\cite{Liu:2026cpf}, on the other hand, presents a package for solving canonical differential equations efficiently using Chebyshev polynomials.
Both approaches rely on \textsc{Mathematica} and provide extensive interfaces in the Wolfram language\footnote{In Ref.~\cite{Baur:2026zlw}, the actual integration of the differential-equation system is performed in C\texttt{++}.}. With Monte Carlo integration of virtual corrections in mind, we aim for a standalone implementation that can evaluate the solution at a very large number of phase-space points and can be parallelized efficiently on computer clusters. For this reason, we avoid any dependence on commercial software at runtime.
Moreover, we note that Ref.~\cite{Liu:2026cpf} uses \textsc{Mathematica} to evaluate elliptic functions appearing in one of its examples, while Ref.~\cite{Baur:2026zlw} implements such functions in C\texttt{++} through series expansions. Our strategy is different: we do not evaluate special functions explicitly. Instead, all geometric building blocks are propagated by their own rational differential equations. This makes the method applicable, at least in principle, to any system of differential equations which has an underlying rational structure, which not only goes beyond the elliptic case, but is also always the case for Feynman integrals. We expect, therefore, our method to be universally applicable for arbitrary  Feynman integrals.

The paper is organized as follows. In the next section we present the method and introduce the necessary concepts. We then apply it to a realistic example containing the complications encountered in current multi-scale calculations, avoiding toy models in order to demonstrate the readiness of the approach for large-scale applications. We subsequently describe the accompanying software package and its use in phenomenological computations. We close with conclusions and an outlook.

\section{Idea}

Consider a set of master integrals $\{ I_i(\vec{x},\epsilon) \}_{i=1}^{n_I}$ depending on parameters $\vec{x} = \big( x_1, \dots, x_{n_x}\big)$ and the dimensional-regularization parameter $\epsilon$. In most cases, $x_i$ are internal masses and scalar products of external momenta. We will use the vector notation, $\vec{I}$, for the master integrals as well. As a consequence of integration-by-parts identities,\footnote{An introduction to the subject can be found in Ref.~\cite{Weinzierl:2022eaz}.} the master integrals satisfy a system of first-order linear differential equations 
\begin{equation}
    \dd \vec{I} = \bm{M}(\vec{x},\epsilon) \vec{I} \,,
\end{equation}
where $\bm{M}(\vec{x},\epsilon)$ is a matrix-valued one-form with rational-function coefficients,
\begin{equation}
    \bm{M}(\vec{x},\epsilon) = \sum_{k=1}^{n_x} M_k(\vec{x},\epsilon) \, \dd x_k \,.
\end{equation}
The matrices $M_k$ define a mapping of the vector of master integrals, $I_i \to \sum_j \big( M_k \big)_{ij} I_j$. Suppose there exists an invertible matrix $U \equiv U(\vec{x},\epsilon)$ such that
\begin{equation} \label{eq:rotation}
    U \bm{M}(\vec{x},\epsilon) U^{-1} - U \dd U^{-1} = \epsilon \bm{A}(\vec{x}) \,.
\end{equation}
Then
\begin{equation} \label{eq:canonical-diffeqs}
    \dd \vec{J} = \epsilon \bm{A}(\vec{x}) \vec{J} \,, \qquad \vec{J} \equiv U \vec{I} \,.
\end{equation}
This is Henn's canonical differential equation \cite{Henn:2013pwa} with connection matrix $\bm{A}(\vec{x})$ and canonical master integrals $\vec{J}$ (canonical basis). The matrix $U$ and the connection matrix $\bm{A}$ may both depend on $\vec{x}$ through algebraic and transcendental functions $f_i(\vec{x})$, $i = 1,\dots,n_f$. It is always possible to include factors of $\epsilon^n$, $n \in \mathbb{N}$ in $U$ so that the canonical master integrals admit a Taylor expansion in $\epsilon$. The objective is to obtain the coefficients of the Taylor expansion up to a predefined order at point $\vec{x}_1$ by numerically solving Eq.~\eqref{eq:canonical-diffeqs} with a boundary condition at a non-singular point $\vec{x}_0$.

Direct numerical solution of differential equations for master integrals has been shown to be an efficient method already in Refs.~\cite{Boughezal:2007ny,Czakon:2008zk}, long before the canonical form was proposed. Recently, the method has been applied in cases where the differential equation, albeit not fully canonical, already contains square roots. In Ref.~\cite{PetitRosas:2025xhm}, the analytic continuation of square roots has been completely avoided by choosing appropriate integration paths. On the other hand, in Ref.~\cite{Becchetti:2025rrz}, an algorithm for the analytic continuation of square roots was presented that allows to choose arbitrary integration paths. These publications demonstrate that the presence of functions other than rational is a challenge in practice. Still, a canonical differential equation is substantially more compact than the differential equation for arbitrary master integrals. If there exists a stable, efficient and smooth implementation of the occurring special functions, then one may expect a more stable and efficient solution of the differential equation. What if, however, the required implementation is not available? Is it worth investing the time to implement the special functions on a project-by-project basis while building a database for future applications?

Our idea is to treat the problem as a non-linear first-order differential-equation system consisting of the canonical differential equation \eqref{eq:canonical-diffeqs} and a differential equation for the special functions,
\begin{equation} \label{eq:special-functions-diffeqs}
    \dd f_i = V_i(\vec{x},\{ f_j \}) \,.
\end{equation}
The derivatives of the special functions, gathered in the vector field $V_i$, must be known anyway in order to obtain the connection matrix according to Eq.~\eqref{eq:rotation}. For a square root $r \equiv \sqrt{g(\vec{x})}$, for example, one has
\begin{equation}
    \dd r = \frac{r}{2 g} \dd g \,.
\end{equation}
If any properties of the special functions, e.g.\ algebraic relations, have been used in order to simplify the connection matrix, then the boundary values of the special functions at $\vec{x}_0$ must also satisfy these properties. Otherwise, the boundary values are entirely arbitrary as far as the structure of the canonical differential equation is concerned.\footnote{As an obvious example, in the Calabi-Yau case, one has to pay attention that the boundary values chosen for the periods are consistent with the form of the Griffith's relations one has used to simplify the connection matrix.} On the other hand, it is important to provide the correct condition for the canonical master integrals. This can either be given in analytic form, if known, or be determined numerically, for example by calculating the master integrals $\vec{I}$ at $\vec{x}_0$ in a Laurent expansion in $\epsilon$ to sufficient order using \textsc{AMFlow} \cite{Liu:2022chg}. These master integrals are then to be rotated to the canonical basis using the matrix $U$ at $\vec{x}_0$ with the boundary values of the special functions as described above. The integration from $\vec{x}_0$ to $\vec{x}_1$ proceeds along a complexified path in order to avoid singularities of the connection matrix. The correct analytic continuation of the special functions is certified by the use of the differential equation and does not require any further attention. We have verified that this approach applied to integrals of Ref.~\cite{Becchetti:2025rrz} is only marginally less efficient than using the complicated algorithm for the analytic continuation of square roots presented there.

\section{Example} \label{sec:example}

In order to illustrate the method, we consider the 220 two-loop master integrals occurring in the amplitudes for di-$\gamma$, di-jet and $\gamma$+jet hadro-production through a virtual loop of quarks of mass $m \neq 0$ \cite{Coro:2025vgn}. These master integrals are obtained by extending the set of 165 integrals previously considered for the di-photon final state \cite{Becchetti:2025rrz} by a subset of these 165 integrals with some of the external lines crossed. Although 11 of the integrals are related linearly to the remaining 209, we consider the complete set of 220 just as in Ref.~\cite{Coro:2025vgn}. After Laurent expansion in $\epsilon$ of the canonical-basis integrals, there are $5 \times 220$ functions that need to be evaluated in order to obtain the infrared finite remainder of the renormalized amplitudes.\footnote{Details of the infrared and ultraviolet renormalization procedures that define the finite remainders of the amplitudes can be found in Ref.~\cite{Coro:2025vgn} and are of no relevance to the present publication.} By construction, the functions corresponding to the lowest power of $\epsilon$ are constant. Hence, it is only necessary to solve the canonical differential equation for 880 functions. The connection matrix of the canonical differential equation depends on the standard Mandelstam variables $s,t$ and $u=-s-t$, square roots,
\begin{alignat*}{3}
r_1 &= \sqrt{s(s-4m^2)}\,, \quad &&r_2 &&= \sqrt{t(t-4m^2)}\,, \\ 
r_3 &= \sqrt{(s + t)(4m^2 + s + t)}\,, \quad &&r_4 &&=\sqrt{s(s+4m^2)}\,, \\
r_5 &= \sqrt{t(t+ 4m^2)}\,, \quad &&r_6 &&= \sqrt{(s + t)( s + t - 4m^2)}\,, \\ 
r_7 &= \sqrt{st(st - 4m^2(s + t))}\,, \quad &&r_8 &&= \sqrt{s(s + t)(s(s + t)-4m^2t )}\,, \\
r_9 &= \sqrt{t(s + t)(t(s + t)-4m^2s )}\,, \quad &&r_{10} &&= \sqrt{s(-4m^2(s + t)^2 + s(m^2 + s + t)^2)}\,, \\
r_{11} &= \sqrt{s(-4m^2t^2 + s(-m^2 + t)^2)}\,, \quad &&r_{12} &&= \sqrt{(s + t)(4m^2t^2 + (-m^2 + t)^2(s + t))}\,, \\
r_{13} &= \sqrt{t(-4m^2s^2 + (-m^2 + s)^2t)}\,, \quad &&r_{14} &&= \sqrt{(s + t)(4m^2s^2 + (-m^2 + s)^2(s + t))}\,, \\
r_{15} &= \sqrt{t(-4m^2(s + t)^2 + t(m^2 + s + t)^2)}\,, \quad &&r_{16} &&= \sqrt{-(m^2st(s + t))} \,,
\end{alignat*}
and two types of transcendental functions. The first one has an elegant interpretation as the period of the elliptic curve $E : X \to Y$,
\begin{equation}
    Y^2 = P_4(X)\,, \quad P_4(X) = (m^2-X) (m^2+s-X) \left(m^2 (m^2-3 s)-X (2 m^2+s)+X^2\right)\,. \label{eq:ellcurve}
\end{equation}
With the normalization adopted in Ref.~\cite{Coro:2025vgn}, the period is given by
\begin{equation}
    \varpi^{[0]}_0 (s,m^2) = \frac{1}{\pi} \int_{\mathcal{C}} \frac{\mathrm{d}X}{\sqrt{P_4(X)}} \,,
\end{equation}
where $\mathcal{C}$ is any complex contour enclosing only the branch points $X = m^2$ and $X = m^2 + s$ of the square root. The square root itself is analytically continued along $\mathcal{C}$ and hence smooth. The superscript $[0]$ singles out the period that is regular at $s/m^2 = 0$, which implies the choice of the contour. The integral can be expressed through an elliptic integral of the first kind,
\begin{equation}
    \varpi^{[0]}_0 (s,m^2) = \frac{2}{\pi \sqrt{s m^2}} K\bigg( -\frac{s}{16m^2} \bigg) \,.
\end{equation}
The second type of transcendental functions is defined as follows.
\begin{equation}
    G(s,t,m^2) = \int^{m^2} \mathrm{d}x \, \frac{s (s+2 t)\sqrt{\eval{P_4(x - t)}}_{m^2 = x}}{(t (s+t)-4 s x)^2} \, \varpi_0 (s,x) \,. \label{eq:Gdef}
\end{equation}
Up to an integration constant, which could be determined by matching the expansion in Eq.~(4.34) of Ref.~\cite{Becchetti:2025rrz}, the function can be expressed through elliptic functions of the first and third kind,
\begin{equation}
    G(s,t,m^2) = \frac{4(s+2t)\sqrt{s m^2}}{\pi r_9} \bigg( \frac{x}{y} K(x) + \frac{y-x}{y} \Pi(y,x) \bigg) \,, \quad x = -\frac{s}{16m^2} \,, \quad y = -\frac{s^2}{4t(s+t)} \,.
\end{equation}
Since the master integrals cover the permutations of the external massless momenta, the complete set of transcendental functions also contains
\begin{equation}
    \varpi^{[0]}_0 (t,m^2) \,, \quad \varpi^{[0]}_0 (u,m^2) \,, \quad G(t,u,m^2) \,, \qand G(u,s,m^2) \,.
\end{equation}

In order to numerically solve the canonical differential equation, we construct a non-linear system of first-order differential equations that includes the differential equations for the square roots and the transcendental functions. An ancillary file to Ref.~\cite{Coro:2025vgn} provides the derivatives of $\varpi_0$,
\begin{subequations}\label{eq:diff-relations-varpi0}
\begin{align}
\frac{\partial^2 \varpi_0(s,m^2)}{\partial (m^2)^2}
&=
-\frac{4\,\varpi_0(s,m^2)}{m^2\bigl(16m^2+s\bigr)}
-
\frac{32m^2+s}{m^2\bigl(16m^2+s\bigr)}
\frac{\partial \varpi_0(s,m^2)}{\partial m^2} \,,
\label{eq:diffrel-varpi0-mm}
\\[0.5em]
\frac{\partial \varpi_0(s,m^2)}{\partial s}
&=
-\frac{\varpi_0(s,m^2)}{s}
-
\frac{m^2}{s}\frac{\partial \varpi_0(s,m^2)}{\partial m^2} \,,
\label{eq:diffrel-varpi0-s}
\\[0.5em]
\frac{\partial^2 \varpi_0(s,m^2)}{\partial s\,\partial m^2}
&=
\frac{4\,\varpi_0(s,m^2)}{s\bigl(16m^2+s\bigr)}
-
\frac{1}{16m^2+s}\frac{\partial \varpi_0(s,m^2)}{\partial m^2} \,,
\label{eq:diffrel-varpi0-sm}
\end{align}
\end{subequations}
and the derivatives of $G$,
\begin{subequations}\label{eq:diff-relations-G}
\begin{align}
\frac{\partial G(s,t,m^2)}{\partial s}
&=
\frac{m^2\bigl(16m^2s-t(3s+2t)\bigr)\,
      \varpi_0(s,m^2)\,r_9(s,t,m^2)}
     {(s+t)\,\Delta(s,t,m^2)^2}
\nonumber\\
&\quad
-
\frac{m^2\bigl(16m^2+s\bigr)\,r_9(s,t,m^2)}
     {2(s+t)\,\Delta(s,t,m^2)}
\frac{\partial \varpi_0(s,m^2)}{\partial m^2}\,,
\label{eq:diffrel-G-s}
\\[0.75em]
\frac{\partial G(s,t,m^2)}{\partial t}
&=
-
\frac{m^2s^2\bigl(16m^2+s\bigr)\,
      \varpi_0(s,m^2)\,r_9(s,t,m^2)}
     {t(s+t)\,\Delta(s,t,m^2)^2}
\nonumber\\
&\quad
+
\frac{m^2s\bigl(16m^2+s\bigr)\,r_9(s,t,m^2)}
     {2t(s+t)\,\Delta(s,t,m^2)}
\frac{\partial \varpi_0(s,m^2)}{\partial m^2}\,,
\label{eq:diffrel-G-t}
\\[0.75em]
\frac{\partial G(s,t,m^2)}{\partial m^2}
&=
\frac{s(s+2t)\,\varpi_0(s,m^2)\,r_9(s,t,m^2)}
     {\Delta(s,t,m^2)^2} \,,
\label{eq:diffrel-G-m}
\end{align}
\end{subequations}
where
\begin{equation}
    \Delta(s,t,m^2) = t(s+t)-4m^2s \,.
\label{eq:Delta}
\end{equation}
As independent transcendental functions, we take $\varpi_0(s,m^2)$, $\pdv*{m^2} \varpi_0(s,m^2)$, $G(s,t,m^2)$ and their versions with cyclic permutations of $s,t,u$. The complete system of differential equations obtained in this way relies on the algebraic definition of the square roots,
\begin{equation}
    \big( \sqrt{x} \big)^2 = x \,.
\end{equation}
Hence, the boundary values of the square roots must be chosen to satisfy this constraint, although the branch is irrelevant and we simply choose the principal one. On the other hand, the system does not depend on any other property of the transcendental functions, except for the differential equations \eqref{eq:diff-relations-varpi0}~and~\eqref{eq:diff-relations-G}. In particular, it neither depends on the normalization nor on the regularity of $\varpi_0$ at any point, nor on the integration constant of $G$. Hence, the boundary conditions for the transcendental functions are entirely arbitrary, and we set them to 1 for each of the functions at the boundary point. In principle, the homogeneity of the amplitudes allows us also to set $m^2 = 1$ without loss of generality. We do not make such a substitution in the system. We have checked that this substitution does not substantially reduce the evaluation time of the connection matrix. On the other hand, our phase space points are chosen with $m^2 = 1$.

We consider the following three phase space points with $\vec{x} \equiv (s,t,m^2)$.
\begin{equation}
\vec{x}_1 = \bigg( \frac{11}{3}, -\frac{11}{10}, 1 \bigg) \,, \qquad
\vec{x}_2 = \bigg( \frac{1}{40}, -\frac{1249}{50000}, 1 \bigg) \,, \qquad
\vec{x}_3 = \bigg( 8, -\frac{37}{10}, 1 \bigg) \,.
\end{equation}
At each of these points, we evaluate the master integrals with \textsc{AMFlow}. We take the values of the canonical-basis functions at $x_1$ as the initial condition for the solution of the differential equations. The translation between the original master integrals evaluated with \textsc{AMFlow} and the canonical-basis integrals is performed with the help of the definitions in an ancillary file to Ref.~\cite{Coro:2025vgn} and the boundary values of the square roots and transcendental functions discussed above. We then integrate to $x_i$, $i=2,3$ along the paths $\gamma_i : [0,1] \ni t \to \vec{z}_i(t)$,
\begin{equation} \label{eq:deformation}
    z_{ik}(t) = x_{1k} + (t + 4i\delta_k t(1-t)) (x_{ik} - x_{1k} ) \,.
\end{equation}
The complex-deformation parameters are set to $\vec{\delta} = (0.1,0.2,0)$. The details of the software are provided in the next section. Here, we only comment on the performance as summarized in Tab.~\ref{tab:performance-comparison}.

\begin{table}[htbp]
\centering
\caption{Performance comparison for different hardware and floating-point types. The code has been compiled with \texttt{g++-15 -O2 -DNDEBUG -DBOOST\_UBLAS\_NDEBUG}. \texttt{dd\_real} is the quadruple-precision type of the \texttt{QD} library \cite{QD}. For further details, see text.}
\label{tab:performance-comparison}
\small
\setlength{\tabcolsep}{4pt}
\begin{adjustbox}{max width=\textwidth}
\begin{tabular}{lrrrrrr}
\toprule
& \multicolumn{2}{c}{Apple M4 Pro}
& \multicolumn{4}{c}{AMD Ryzen 5 2400G} \\
\cmidrule(lr){2-3}\cmidrule(lr){4-7}
Point & $x_2$ & $x_3$ & & $x_2$ & & $x_3$ \\
Float type & \texttt{double} & \texttt{double} & \texttt{double} & \texttt{dd\_real} & \texttt{dd\_real} & \texttt{double} \\
Error requested & $10^{-8}$ & $10^{-10}$ & $10^{-8}$ & $10^{-10}$ & $10^{-20}$ & $10^{-10}$ \\
\midrule
Number of steps & 34 & 28 & 35 & 40 & 97 & 28 \\
Number of derivative evaluations & 2199 & 1509 & 2281 & 2984 & 7927 & 1509 \\
Connection-matrix evaluation time [ms] & 0.03 & 0.03 & 0.16 & 2.05 & 2.04 & 0.16 \\
Total time [s] & 0.10 & 0.05 & 0.43 & 7.51 & 19.87 & 0.29 \\
\midrule
Actual max absolute error & \num{1e-6} & \num{6e-10} & \num{1e-6} & \num{7e-9} & \num{8e-20} & \num{6e-10} \\
\bottomrule
\end{tabular}
\end{adjustbox}
\end{table}

As we discuss in the next section, while integrating the system of differential equations, we recommend to aim for a predefined relative error of the actual amplitude w.r.t.\ the Born amplitude. Nevertheless, in order to keep the example software small, we aim here at a predefined maximal absolute error of the evaluated functions (canonical-basis functions and special functions). Since there is $m^2 = 1$, $0.1 < s,|t| < 10$ and the phase space points are not very close to any singularities, the values of the functions are not scattered over orders of magnitude to an extent that would require the use of relative errors. We request a local error of $10^{-8}$ for $x_2$ and $10^{-10}$ for $x_3$ in double precision. It turns out that requesting a lower local error for $x_2$ does not lead to more precise results due to numerical instabilities encountered in double precision. We also solve the system in quadruple precision for $x_2$ with local errors of $10^{-10}$ and $10^{-20}$. The comparison with the master integrals obtained with \textsc{AMFlow} is performed by translating the results of \textsc{AMFlow} at $\vec{x}_{2,3}$ to the canonical basis using the special-function values obtained from the numerical solution of the system of differential equations. We observe that the actual error is $\order{10^{-6}}$ for $x_2$ and $\order{10^{-9}}$ for $x_3$ in double precision. When working in quadruple precision, we achieve an error of $\order{10^{-8}}$ for $10^{-10}$ requested, and of $\order{10^{-19}}$ for $10^{-20}$ requested. The quadruple-precision version is about $10 \times$ slower than the double-precision version as far as the evaluation time of the connection matrix is concerned. We also observe the drastically better performance of Apple Silicon chips compared to AMD. Finally, we point out that we have not performed any special optimization of the connection matrix, at variance with our recommendations in the next section.

We provide the software for the example in the ancillary files to the present publication. The package contains pre-generated code for the connection matrix, the differential equations for the special functions, a file containing the boundary condition at $\vec{x}_1$, and high-precision numerical values obtained at $\vec{x}_{2,3}$. The example program \texttt{test} checks whether the results match the precomputed values within the requested error multiplied with a factor of one hundred. The \textsc{Mathematica} notebook used to generate the code is also provided, although not needed to run the example. On the other hand, if one would like to perform a comparison between the \textsc{AMFlow} values and the numerical solution, then it is necessary to run the second provided notebook. The package should be sufficient to easily produce code for a different application.

\section{Software}

Besides presenting our idea and illustrating it with an example, we would also like to provide a lightweight tool that can be included in a software for the calculation of scattering amplitudes based on canonical differential equations for the master integrals. This should make it easier to obtain novel higher-order predictions for scattering processes. To this end, we have prepared an interface to a tried-and-tested differential-equation solver. This interface highlights all the features discussed in the previous sections. It does not rely on any expensive commercial software, as it is written in C\texttt{++}, and can be compiled with any modern compiler, e.g.\ \texttt{g++} or \texttt{clang}. Furthermore, it only requires the standard \textsc{Boost} library that is usually shipped with the compilers. The actual differential-equation solver is \code{boost::numeric::odeint::bulirsch_stoer} as used first in Ref.~\cite{Czakon:2020vql} and recently for more complicated cases in Refs.~\cite{PetitRosas:2025xhm,Coro:2025vgn}. The code is also prepared for higher floating-point precision through the \texttt{QD} \cite{QD} library that relies on the Intel \texttt{x86\_64} architecture. Other floating-point implementations can also be used thanks to templated types. The software package is contained in the ancillary files to the present publication.

\paragraph{Interface.}
The public interface consists of the class template \code{diffeqs<Real>}, a type bundle \\ \code{diffeqs_types<Real>}, and a small hierarchy of solver-specific exceptions.  The template parameter \code{Real} fixes the real arithmetic type; by default it is \code{double}.  Complex numbers are represented by \code{std::complex<Real>}, while complex vectors and sparse matrices are represented by Boost uBLAS containers.

\paragraph{Fundamental types.}
For a fixed real type \code{Real}, the type bundle \code{diffeqs_types<Real>} defines the aliases listed in Table~\ref{tab:diffeqs-types}.  The solver class \code{diffeqs<Real>} re-exports the same aliases as public member types.

\begin{table}[htbp]
\centering
\small
\begin{tabularx}{\linewidth}{@{}lX@{}}
\toprule
Alias & Meaning \\
\midrule
\code{real} & The scalar real type, equal to \code{Real}. \\
\code{cplx} & The complex scalar type \code{std::complex<Real>}. \\
\code{real_vec} & A real vector, implemented as \code{std::vector<Real>}. \\
\code{cplx_vec} & A complex vector, implemented as \code{ublas::vector<cplx>}. \\
\code{cplx_vec_range} & A view into a complex vector, implemented as \code{ublas::vector_range<cplx_vec>}. \\
\code{matrix} & A sparse complex matrix, implemented as \code{ublas::compressed_matrix<cplx>}. \\
\bottomrule
\end{tabularx}
\caption{Scalar, vector, and matrix types used by the solver interface.}
\label{tab:diffeqs-types}
\end{table}

\paragraph{Types for user-defined functions.}
The problem-specific parts of the system are supplied through callback objects.  Their signatures are
\begin{lstlisting}[style=cppapi]
using connection_t = std::function<void(
    const cplx_vec& x, const cplx_vec& dx,
    const cplx_vec_range& f, matrix& A)>;

using vector_field_t = std::function<void(
    const cplx_vec& x, const cplx_vec& dx,
    const cplx_vec_range& f, cplx_vec_range& df)>;

using path_t = std::function<void(
    const real_vec& x0, const real_vec& x1, real t,
    real_vec& x, real_vec& dx)>;

using error_estimator_t = std::function<real(
    const cplx_vec& f, const cplx_vec& f_err)>;
\end{lstlisting}
The callback \code{connection_t} fills the connection matrix \code{A} of the canonical system at the current complex point \code{x}, tangent \code{dx} (i.e., $\dv*{\vec{x}}{t}$), and current special-function values \code{f}.  The callback \code{vector_field_t} fills the derivatives \code{df} (i.e., $\dv*{ f_i}{t}$) of the special functions. A \code{path_t} object supplies the point \code{x} and tangent \code{dx} at \code{t} on the integration path $\vec{x}(t)$, subject to constraints \code{x = x0} at \code{t = 0} and \code{x = x1} at \code{t = 1}.  Finally, \code{error_estimator_t} maps the function values \code{f} and their errors \code{f_err} to the local-error estimate used by the solver.

\paragraph{Auxiliary arithmetic.}
The namespace \code{diffeqs_int_cplx} provides arithmetic overloads between integral types and \code{std::complex<Real>}. These overloads are independent of the solver class itself and are a convenience for writing expressions that mix integer constants with complex quantities, which typically occur in the connection-matrix code.

\paragraph{Exceptions.}
All solver-specific exceptions derive from \code{diffeqs_exception}, which itself derives from \code{std::runtime_error}. The derived exceptions report violations of user-configured stopping criteria, as summarized in Table~\ref{tab:diffeqs-exceptions}.

\begin{table}[htbp]
\centering
\small
\begin{tabularx}{\linewidth}{@{}lX@{}}
\toprule
Exception & Condition \\
\midrule
\code{diffeqs_exception} & Common base class for all solver-specific exceptions. \\
\code{diffeqs_max_steps_exception} & The number of integration steps exceeded the limit set by \code{set_max_steps}. \\
\code{diffeqs_max_evals_exception} & The number of derivative evaluations exceeded the limit set by \code{set_max_evals}. \\
\code{diffeqs_max_time_exception} & The evaluation time exceeded the limit set by \code{set_max_time}. \\
\code{diffeqs_min_step_exception} & The adaptive step size fell below the value set by \code{set_min_step}. \\
\bottomrule
\end{tabularx}
\caption{Solver-specific exceptions.}
\label{tab:diffeqs-exceptions}
\end{table}

\paragraph{Class interface.}
The main entry point is \code{diffeqs<Real>}.  Its public interface is summarized below.

\begin{lstlisting}[style=cppapi]
template<class Real = double>
class diffeqs {
public:
  using real              = typename diffeqs_types<Real>::real;
  using cplx              = typename diffeqs_types<Real>::cplx;
  using real_vec          = typename diffeqs_types<Real>::real_vec;
  using cplx_vec          = typename diffeqs_types<Real>::cplx_vec;
  using cplx_vec_range    = typename diffeqs_types<Real>::cplx_vec_range;
  using matrix            = typename diffeqs_types<Real>::matrix;
  using connection_t      = typename diffeqs_types<Real>::connection_t;
  using vector_field_t    = typename diffeqs_types<Real>::vector_field_t;
  using path_t            = typename diffeqs_types<Real>::path_t;
  using error_estimator_t = typename diffeqs_types<Real>::error_estimator_t;

  diffeqs(connection_t connection,
          vector_field_t vector_field,
          std::istream& boundary);

  void evaluate(const real_vec& x,
                const real_vec& deformation,
                const real& error,
                cplx_vec& f);

  void change_error_estimator(error_estimator_t e);
  void change_path(path_t path);
  void set_log_stream(std::ostream& out);
  void set_max_steps(unsigned n);
  void set_max_evals(unsigned n);
  void set_max_time(unsigned n);
  void set_min_step(real d);
};
\end{lstlisting}

\paragraph{Constructor.}
The constructor receives the connection-matrix callback, the special-function vector field, and a stream containing the boundary data.  

\paragraph{Boundary stream.}
The boundary stream passed to the constructor provides the initial condition for the solution of the differential-equation system. Its expected layout is
\begin{lstlisting}[style=cppapi]
dimx
x_1 ... x_dimx
order dimbasis nspecial
for j = 0,...,order:
  for i = 1,...,dimbasis:
    Re(f_i_j) Im(f_i_j)
for k = 1,...,nspecial:
  Re(special_f_k) Im(special_f_k)
\end{lstlisting}
Here \code{dimx} is the dimension of the kinematic point and \code{x_i} are the components of the boundary point. The number of terms of the Taylor expansion of the canonical master integrals is \code{order+1}, the dimension of the canonical basis is \code{dimbasis} and the number of special functions is \code{nspecial}. \code{f_i_j} is the coefficient of $\epsilon^j$ in the Taylor expansion of the $i$-th canonical master integral, and \code{special_f_k} is the value of the $k$-th special function at the boundary point.

\paragraph{Solution of the differential-equation system.}
A call to \code{evaluate} integrates the system from the boundary point to the real endpoint \code{x}. The vector \code{deformation} defines the complex deformation of the real integration path, \code{error} is the requested accuracy, and the result is written to \code{f}. The function returns normally when the requested evaluation has completed; stopping criteria are reported by the exceptions in Table~\ref{tab:diffeqs-exceptions}.

Let $\vec{x}_0$ denote the boundary point and let $\vec{x}(t)$ be the real path produced by the current \code{path_t} object, with $\vec{x}(0)=\vec{x}_0$ and $\vec{x}(1)=\vec{x}_1$. By default, the real path is a straight line,  $\vec{x}(t)=\vec{x}_0+t(\vec{x}_1-\vec{x}_0)$. During evaluation, the real-path point is mapped to the complex integration point
\begin{equation} \label{eq:path-complexification}
    z_{k}(t) = x_{0k} + (1 + 4i\delta_k (1-t)) (x_{k}(t) - x_{0k} ) \,,
\end{equation}
where $\delta_k$ is the corresponding component of \code{deformation}. For the default real path, the complex integration path matches Eq.~\eqref{eq:deformation}.

The solver attempts to reach the requested \code{error} for a local-error estimate provided by an \\ \code{error_estimator_t} object. The default local-error estimate is the maximal absolute error of the functions estimated by the solver. 

The output vector \code{f} has the same coefficient ordering as the boundary data, but it contains only the non-constant functions. In particular, it omits the Taylor coefficients $f_{i,0}$ and stores
\begin{lstlisting}[style=cppapi]
(f_1_1,f_2_1,...,f_dimbasis_1,f_1_2,...,f_dimbasis_2,...,f_dimbasis_order,special_f_1,...,special_f_nspecial).
\end{lstlisting}

\paragraph{Optional functionality.}
The default real path can be replaced by \code{change_path}. The default local-error estimate can be replaced by \code{change_error_estimator}. Diagnostic output during \code{evaluate} can be enabled by \code{set_log_stream}. The remaining setter functions configure the limits that trigger the corresponding exceptions.

\paragraph{Recommendations.}
Experience from the computations presented in Refs.~\cite{Badger:2025ljy,Badger:2025ilt} leads to the following recommendations concerning the application of the tools.
\begin{enumerate}
    \item {\bf Error estimation during integration.} The default \code{error_estimator_t} object returns the maximal absolute error of the functions estimated by the solver. While this is sufficient for testing and experimentation, it is certainly suboptimal for production runs. Ultimately, the error estimation should be based on the amplitude that is being calculated. If this amplitude is expressed as a vector of coefficients of monomials in the functions, then, using the solver-estimated values and errors, one can easily obtain an estimate of the absolute error on the amplitude by standard error propagation (an example can be found in Section 4 of Ref.~\cite{Badger:2025ljy}). A user-defined routine for this purpose can be set with \code{change_error_estimator}. Although one can aim for an absolute precision of the amplitude being evaluated, it is probably more efficient to rather aim for a relative error w.r.t.\ to the Born amplitude. This should be taken into account in preparing the \code{error_estimator_t} function.
    \item {\bf Error estimation of the result.} A reasonable error estimate of the result for the amplitude (global error) can be obtained by integrating the system from two different boundary points and comparing the outcomes. This was the method used in Ref.~\cite{Badger:2025ljy}. It provides an additional layer of safety against incorrect analytic continuation to an unphysical Riemann sheet as well.
    \item {\bf Integration path and analytic continuation.} The integration path is a complex deformation of a real path defined in Eq.~\eqref{eq:path-complexification}. By default, this real path is a straight line between the boundary point and the evaluation point. The complex deformation is used to avoid numerical instabilities. The parameters of the deformation may be tuned by experimentation, but values of the order of 1/10 seem to work in most cases. A straight-line real path may be used, even if it leaves the physical phase space, since the complex deformation provides a means of analytic continuation. On the other hand, the user might want to stay within the physical phase space. This may be achieved by replacing the path parameterization with \code{change_path}. A path within the physical phase space was used in Ref.~\cite{Badger:2025ljy} by exploiting the fact that phase-space parameterizations are maps from a hypercube (a convex set). Independently, it is recommended to take a boundary point within the physical phase space, not least for efficiency reasons. In case of doubts on the correctness of analytic continuation, integration from different boundary points can be used as a test.
    \item {\bf Stopping conditions.} If the \code{error} requested in a call to \code{evaluate} cannot be reached, then the solver may end in an infinite loop. For this reason a stopping condition should be set using one of the available criteria. The exception that will be thrown should be caught and dealt in one of two ways: either increase the error requested or switch to higher floating-point precision. A reasonable value for the stopping criteria can be found with some experimentation.
    \item {\bf Optimization.} The solution time is dominated by the evaluation of the connection matrix. For this reason, effort should be put into the optimization of the latter. The simplest optimization is to set one of the parameters to a numerical value using dimensional analysis (although this did not significantly improve performance for the example in Section~\ref{sec:example}). In Ref.~\cite{Badger:2025ljy}, an optimization strategy was presented which started with using \textsc{MultivariateApart} \cite{Heller:2021qkz} in order to perform a partial-fraction decomposition of the rational functions. Each partial fraction was then factorized into independent polynomials. The polynomials were simplified to reduce the number of floating-point operations and separated for evaluation before insertion into the connection matrix. We note that this strategy increases the numerical stability of the connection-matrix code, which also reduces the evaluation time because fewer solver steps are needed and the switch to higher floating-point precision is rarer. Finally, we recommend to split the code into smaller routines that only evaluate a fraction of the expressions. This makes compilation much faster and allows to compile large connection matrices which otherwise lead to compiler crashes.
    
\end{enumerate}

\section{Conclusions}
In this paper, we have presented a strategy for the direct numerical solution of canonical $\epsilon$-factorized differential equations whose connection matrices contain algebraic or transcendental functions. 
The main observation is that these functions are not external special functions that require an ad-hoc implementation. 
In fact, since they arise from Feynman integrals which, before the transformation
to a canonical basis, always satisfy differential equations with rational coefficients,
the  building blocks required to reach a canonical basis 
themselves satisfy differential equations with rational coefficients. 
By solving these auxiliary equations together with the canonical
system, the problem is reduced to the numerical solution of an enlarged system
with rational coefficients.

We demonstrated the method on the two-loop master integrals relevant for
di-$\gamma$, di-jet and $\gamma$+jet hadro-production through a heavy-quark loop. 
This non-trivial example involves many of the complications expected in realistic multi-scale
calculations, including several square roots and transcendental functions
related to elliptic periods. The calculation shows that no explicit evaluation
of these functions is required, and their analytic continuation is entirely fixed by the
differential equations they satisfy.

We also provided a lightweight C\texttt{++} implementation of the method,
designed for repeated evaluations at many phase-space points and for use in
parallelized amplitude computations. The example provided illustrates that our approach
is already practical without special optimization of the connection matrix,
while leaving clear room for further improvements in code generation,
expression optimization and error estimation.

We expect that our method will be particularly useful for future applications in which
canonical bases involve periods of elliptic curves, Calabi-Yau manifolds,
higher-genus curves, and, possibly, even different geometries. In such cases, the
ability to avoid closed-form implementations of the underlying special
functions will be a crucial ingredient to turn canonical differential equations beyond polylogarithms
into a directly usable tool for numerical phenomenology.

\section*{Acknowledgments}
L.T. is grateful to Christoph Nega and Fabian Wagner for help with the ancillary files
and for many illuminating discussions on canonical forms.
This work was supported in part by the Deutsche Forschungsgemeinschaft (DFG) through the grant 396021762 - TRR 257: Particle Physics Phenomenology after the Higgs Discovery (M.C.) and the Excellence Cluster ORIGINS EXC-2094-390783311 (L.T.) and in part by the European Union through the European Research Council under the grant agreements 949279 (ERC Starting Grant HighPHun) (L.T.).

\bibliographystyle{JHEP}
\bibliography{main} 

\end{document}